\documentclass[%
 twocolumn,
 amsmath,amssymb,aps,prl,superscriptaddress,nolongbibliography]{revtex4-2}

\usepackage{braket}
\usepackage{graphicx}% Include figure files
\usepackage{dcolumn}% Align table columns on decimal point
\usepackage{hyperref}
\usepackage{bm}% bold math
\usepackage{xcolor}

\begin{document}
%\preprint{APS/123-QED}$^{1,2,3}$, 

%\title{Circular dichroism in attosecond transient absorption spectroscopy of atoms}
%\title{Unifying Linear and Circular Dichroism in Attosecond Transient Absorption}
\title{Attosecond Transient Polarimetry}
% Force line breaks with \\

\author{Nicola Mayer}
\email{nicola.1.mayer@kcl.ac.uk}
\affiliation{Attosecond Quantum Physics Laboratory, Department of Physics, King's College London, Strand, London, WC2R 2LS, United Kingdom}

\author{Kylie Gannan}
\affiliation{Department of Chemistry, University of California, Berkeley, California 94720, USA}
\author{Stephen R Leone}
\affiliation{Department of Chemistry, University of California, Berkeley, California 94720, USA}
\affiliation{Chemical Sciences Division, Lawrence Berkeley National Laboratory, Berkeley, California 94720, USA}
\affiliation{Department of Physics, University of California, Berkeley, California 94720, USA}
\author{Lauren B Drescher}
\email{lauren.drescher@mbi-berlin.de}
\affiliation{Max-Born-Institut, Max-Born-Str. 2A, 12489, Berlin, Germany}
\affiliation{Department of Chemistry, University of California, Berkeley, California 94720, USA}
\affiliation{Chemical Sciences Division, Lawrence Berkeley National Laboratory, Berkeley, California 94720, USA}

\date{\today}

\begin{abstract}
We introduce elliptical-dichroic attosecond transient absorption spectroscopy (eDATAS), an ultrafast attosecond spectroscopy technique in which an atomic gas is probed by elliptically polarized attosecond and near-infrared pulses with an arbitrary angle between the polarization ellipses and a varying time-delay. Using perturbation theory, we show that the transient absorption signal factorizes into form factors whose dependence on pulse ellipticities and polarization-ellipse orientations are governed entirely by dipole selection rules: ellipticity sets the amplitudes of the interfering multiphoton pathways, while the relative orientations of the ellipses control the phase. We reinterpret circular- and linear-dichroic attosecond transient absorption spectroscopy as limiting cases of eDATAS, and validate the theory through an eDATAS measurement in helium, finding excellent agreement. Finally, we demonstrate that eDATAS can also serve as an in situ measurement of the polarization state of the XUV field, avoiding the need for a dedicated XUV polarimeter.
\end{abstract}

\maketitle

The exceptional temporal resolution of attosecond extreme-ultraviolet (XUV) pulses makes it possible to probe ultrafast electron dynamics as they unfold on their natural timescale in atomic, molecular, and solid-state systems \cite{Leone:2014aa,Krausz:2009aa,Borrego-Varillas:2022aa}. Combined with the ability to control both the spin and orbital angular momentum degrees of freedom of XUV radiation through High-Harmonic Generation (HHG) \cite{Bengs:2021aa,Hickstein:2015aa,Kfir:2015aa,Gauthier:2017aa} and advanced polarizers \cite{Schmising:2017aa}, attosecond spectroscopy can now investigate circular- and vortex-dichroic phenomena, along with other symmetry-breaking effects driven by electron dynamics triggered by tailored laser pulses. This has led to the development of methods such as circular-dichroic photoelectron spectroscopy \cite{Han:2023aa,Han:2023ab,Serov:2026aa}, magnetic circular dichroic spectroscopy \cite{Siegrist:2019aa,Willems:2015aa,Geneaux:2024aa}, and ultrafast chiroptical spectroscopy \cite{Ayuso:2022aa,Han:2025aa}.

\begin{figure*}
    \centering
    \includegraphics[width=0.75\linewidth]{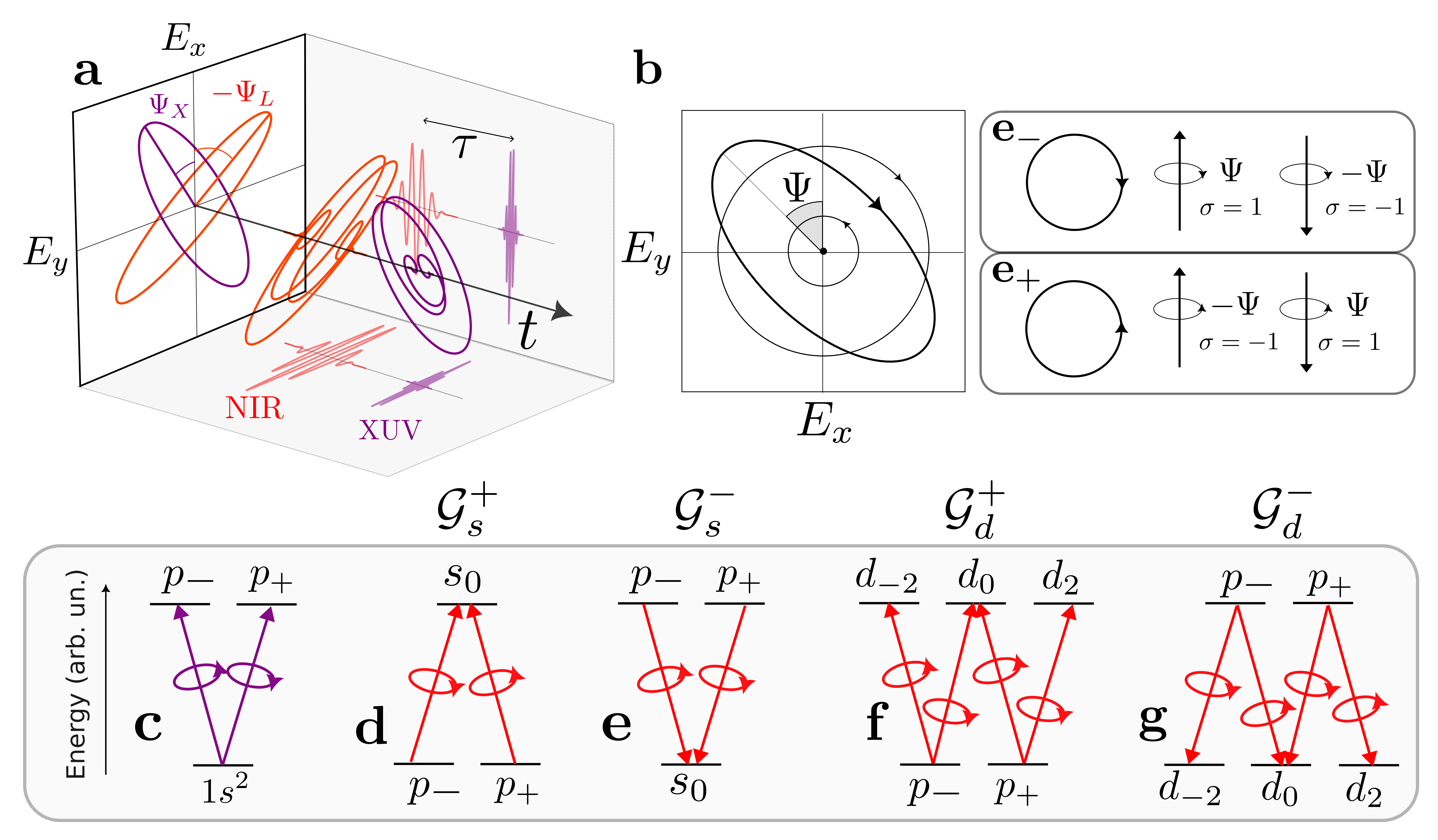}
    \caption{(a) Schematic representation of the XUV (purple) and NIR (red) pulses used in eDATAS. $\tau$ indicates the delay between the peak maximum of the XUV and NIR envelopes, $\Psi_X$ and $\Psi_L$ the angles formed by the major axis of the polarization ellipse with the $y$-axis of the laboratory frame, where angles are measured in the counter-clockwise sense. (b) Schematic description of the definition of elliptically polarized light used in the manuscript. Clockwise $\mathbf{e}_+$ (LCP) polarization carries SAM $\sigma=-1$, while counter-clockwise $\mathbf{e}_-$ (RCP) polarization carries SAM $\sigma=1$. Absorption (emission) of a RCP photon carries a phase delay of $\Psi$ (-$\Psi$), while for a LCP photon absorption (emission) of a photon carries a phase delay of $-\Psi$ ($\Psi$).  (c) The helium ground state $|1s^2\rangle$ is coupled to the $m=\pm1$ ($|p_\pm\rangle$) levels via the absorption of a RCP/LCP XUV photon. (d) The $|p_\pm\rangle$ levels are coupled to a $m=0$ $s$-state ($|s_0\rangle$) with higher energy. (e) Coupling to a $|s_0\rangle$ state with lower energy than the $p_{\pm}$ states.(f) Coupling to the $m=-2,0,2$ magnetic levels of a $d$-state ($|d_{-2}\rangle,|d_0\rangle,|d_2\rangle$ respectively) with higher energy than $p_{\pm}$ states. (g) Coupling to a $d$-state with lower energy than the $p_{\pm}$ states. Above each of the four cases (d-g) we indicate the corresponding form factor $G_{s/d}^{\pm}$ reported in Eqs. \ref{eq:formfactor}.}
    \label{fig:Fig1}
\end{figure*}

Following on this trend we recently introduced a new attosecond technique named circular-dichroic attosecond transient absorption spectroscopy (cDATAS) \cite{Drescher2025aa}. cDATAS combines circularly polarized (CP) XUV attosecond and near-infrared (NIR) few-cycle pulses in an attosecond transient absorption experimental setup, accessing magnetic-quantum-number-dependent couplings while preserving the high spectral and temporal resolution characteristic of ATAS \cite{Wu:2016aa}. cDATAS promises access to spin-dependent electron dynamics via optical orientation of the excited states \cite{Meier:2012aa,Skrotsky:1961aa,Franzen:1957aa}, complementing linear-dichroic ATAS (lDATAS) \cite{Reduzzi:2015aa,Wu:2016aa,Gannan:2025aa}, where configurations with parallel and orthogonal polarizations between XUV and NIR were investigated.
Here, we bridge the gap between these two techniques in atomic media by establishing elliptical dichroic ATAS (eDATAS), showing that both cDATAS and lDATAS can be understood within the more general framework of ATAS driven by elliptically polarized XUV and NIR pulses with an arbitrary relative orientation of their polarization ellipses (see Fig. \ref{fig:Fig1}a).

To establish the technique both experimentally and theoretically, we focus on the ATAS signal in helium, a prototypical target in attosecond experiments \cite{Wu:2016aa,Drescher2025aa,Reduzzi:2015aa,Chini:2012aa}. Using analytical expressions derived from perturbation theory, we first provide an intuitive picture of the dynamics underlying the ellipticity and orientation angle dependence of the ATAS signal. We then record the experimental ATAS signal at a fixed time delay while varying both XUV and NIR pulse ellipticity and orientation in a controlled manner, showing excellent agreement between experiment and theory. Finally, we discuss the use of eDATAS as an in-situ probe of the polarization state of XUV radiation, without the requirement of an additional polarimetry setup operating in the XUV range and show how the characteristic dynamical dependence on the ellipticity and orientation allows to identify the orbital character and relative energetic ordering of NIR coupled states in the atomic target.

Whereas cDATAS~\cite{Drescher2025aa} exploits selection and propensity rules to limit excitation pathways of either co- or counter-rotating circularly polarized pulses to tailor the ATAS response of an atomic gas, the presented eDATAS is more general: All pathways are considered, involving the absorption and/or emission of right- and left-circularly polarized photons. These pathways interfere in the spectral response, producing a strongly modulated signal in which the pulse ellipticities (balance of left- to right-circularly polarized photons) govern the amplitude of the interfering pathways, while the relative orientation of the polarization ellipses controls their phase.

To show this, let us define an elliptically polarized pulse in the circular basis and atomic units as
\begin{eqnarray}
    \label{eq:fields}
    \mathbf{E}(t)&=&\frac{\mathcal{E}_0f(t)}{\sqrt{2(1+\epsilon^2)}}\left[\frac{(1+\epsilon)}{2}e^{\text{i}\Psi}\mathbf{e}_-\right.\nonumber\\
    &&\left.+\frac{(1-\epsilon)}{2}e^{-\text{i}\Psi}\mathbf{e}_+\right]e^{-\text{i}\omega t}+\text{c.c.},
\end{eqnarray}
where $\mathcal{E}_0$ is the electric field strength, $f(t)$ is the envelope of the pulse, $\omega$ is the central frequency of the pulse, \textcolor{green}{$\epsilon$ the ellipticity} and $\mathbf{e}_{\pm}=\mp(\hat{\mathbf{x}}\pm\text{i}\hat{\mathbf{y}})/\sqrt{2}$, where $\mathbf{e}_{\pm}$ correspond respectively to left-/right-circularly polarized (LCP/RCP) light, or counter-clockwise/clockwise rotation as seen from the point of view of the receiver. LCP/RCP photons carry $\hbar/-\hbar$ quanta of angular momentum, respectively, and $\epsilon=\mp1$ corresponds to a LCP/RCP pulse, respectively. Using this definition, the major axis of the polarization ellipse forms an angle $\Psi$ with the $y$-axis of the laboratory (measured in the counter-clockwise sense), as shown in Fig. \ref{fig:Fig1}a.

We denote a state belonging to the $1sn\ell$ Rydberg series of helium as $n\ell_m$, where $n$ is the principal quantum number, $\ell$ is the orbital angular momentum quantum number ($s,p,d$ correspond respectively to states with angular momentum $\ell=0,1,2$) and $m\in[-\ell,\ell]$ is the magnetic quantum number. We choose the quantization axis $z$ along the propagation direction of the fields, and work within the electric dipole approximation by ignoring magnetic dipole and higher multipole interactions with the external field. With these assumptions, the $np_m$ states with $m=\pm1$ are coupled to the helium ground state $1s^2$ via the absorption of RCP/LCP photons, respectively (panel (c) of Fig. \ref{fig:Fig1}). The neighboring Rydberg series $ns$ ($\ell=0$) and $nd$ ($\ell=2$) are coupled to the $np$ states via the absorption/emission of an additional NIR photon. Retaining only one photon transitions, we can then classify four relevant cases in one-photon absorption or emission from a $np$ state, as shown in the panels (d-g) of Fig. \ref{fig:Fig1}. The $1snp$ is either coupled to a $s$-state ($\ell=0$) with energy above the $np$ state (panel (d)) or below the $np$ state (panel (e)), or to a $d$-state ($\ell=2)$ with energy above the $np$ state (panel (f)) or below the $np$ state (panel (g)).

\begin{figure*}
    \centering
    \includegraphics[width=0.75\linewidth]{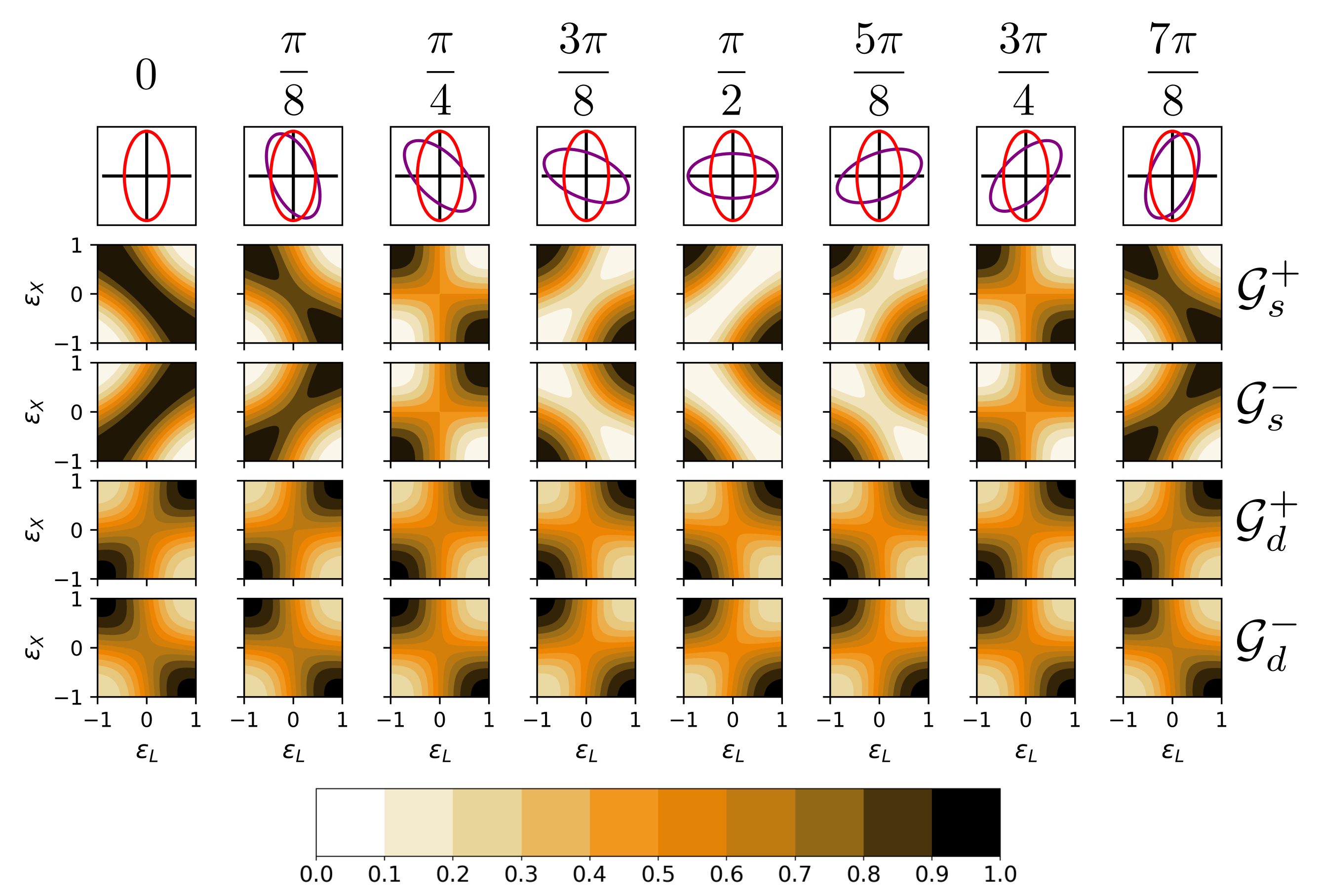}
    \caption{The ellipticity and polarization angle dependence of the eDATAS signal obtained from perturbation theory. The top row shows the polarization ellipses of the XUV (purple) and NIR (red) fields for a varying angle between the polarization ellipses. The second to fifth row correspond to the four cases considered in Fig. 1: $\mathcal{G}_s^+$, $\mathcal{G}_s^-$, $\mathcal{G}_d^+$ and $\mathcal{G}_d^-$. For each map, the x-axis corresponds to the ellipticity of the NIR pulse $\epsilon_L$, and the y-axis to the ellipticity for the XUV pulse $\epsilon_X$, at a fixed angle between the polarization ellipses. As can be seen, the ellipticity dependence is mirrored going from $\mathcal{G}^+$ to $\mathcal{G}^-$. While the $\mathcal{G}_s$ factors show a strong dependence on the polarization angle, leading to a strong suppression of the signal for low ellipticities and orthogonal orientations, the $\mathcal{G}_d$ factors are only weakly dependent. Each map is normalized to its peak maximum.}
    \label{fig:Fig2}
\end{figure*}
For each of these cases, we use perturbation theory (PT) to derive the general dependence of the ATAS response in terms of ellipticities and relative orientation of the polarization ellipses of the two-color field, as derived in Appendix A. In particular we find that while the eDATAS signal in general depends on both XUV frequency $\omega$ and time-delay $\tau$, its ellipticity and orientation angle dependence can be factored out as a form factor  $\mathcal{G}^\alpha_\ell$ for each case of panels (d-g) in Fig. \ref{fig:Fig1}, where $\ell=\{s,d\}$ indicates the angular momentum of the dark state and $\alpha=\pm$ indicates whether the dark state has energy above ($+$) or below ($-$) the $np$ state. These form factors are given by
\begin{widetext}
\begin{eqnarray}
    \label{eq:formfactor}
    \mathcal{G}_s^{\pm}&=&\frac{(1+\epsilon_X)^2(1\mp\epsilon_L)^2+(1-\epsilon_X)^2(1\pm\epsilon_L)^2+2(1-\epsilon_X^2)(1-\epsilon_L^2)\cos(2\Delta\Psi)}{(1+\epsilon^2_X)(1+\epsilon^2_L)},\\
    \mathcal{G}_d^\pm&=&\mathcal{C}_{0}\frac{(1+\epsilon_X)^2(1\mp\epsilon_L)^2+(1-\epsilon_X)^2(1\pm\epsilon_L)^2+2(1-\epsilon_L^2)(1-\epsilon_X^2)\cos(2\Delta\Psi)}{(1+\epsilon^2_X)(1+\epsilon^2_L)}+\nonumber\\
    &&\mathcal{C}_{2}\frac{(1+\epsilon_X)^2(1\pm\epsilon_L)^2+(1-\epsilon_X)^2(1\mp\epsilon_L)^2}{(1+\epsilon^2_X)(1+\epsilon^2_L)}.
\end{eqnarray}
\end{widetext}
Here the subscripts $X/L$ indicate the XUV or NIR field, respectively, $\Delta\Psi=\Psi_X-\Psi_L$ is the relative angle between the major axis of the polarization ellipses, $\epsilon_i$ is the ellipticity of the given pulse, and $\mathcal{C}_0=\frac{1}{30}$, $\mathcal{C}_2=\frac{1}{5}$ are angular factors weighting the transitions to the $m=0$ and $m=\pm2$ sublevels of the $d$ state.

The physical interpretation of these form factors is straightforward. The polynomial terms in the numerators encode the photon absorption and emission pathways, governed by the selection rules. Specifically, terms of the form $(1\pm\epsilon_X)^2(1\mp\epsilon_L)^2$ arise from the absorption of counter-rotating XUV and NIR photons, while $(1\pm\epsilon_X)^2(1\pm\epsilon_L)^2$ arise from the absorption of co-rotating XUV and NIR photons. Which combination contributes to the ATAS signal depends on both the energy of the dark $ns$ or $nd$ state and its magnetic quantum number $m$. 

For example, selection rules require that excitation of an $s$-state ($m=0$) with energy below the $np$ line can occur only via absorption and emission of co-rotating XUV and NIR photons, while an $s$-state ($m=0$) with energy above the $np$ line can only occur via absorption of counter-rotating XUV and NIR photons. This leads to an ATAS signal that maximizes for circularly polarized pulses when $\epsilon_X=\epsilon_L$ if the $s$-state has an energy below the $np$ line and when $\epsilon_X=-\epsilon_L$ if it is above. This is clearly shown in Fig. \ref{fig:Fig2}, where we report the dependence of the $\mathcal{G}_{\ell}^{\pm}$ factors with respect to both the ellipticities and polarization angles. For $\mathcal{G}_s^{+}$ (second row in Fig. \ref{fig:Fig2}), the ATAS signal is maximum for counter-rotating XUV and NIR circularly polarized pulses, while for $\mathcal{G}_s^{-}$ (third row), the signal maximizes for co-rotating pulses. Similarly, for $d$ states represented by $\mathcal{G}_{d}^{\pm}$ (fourth and fifth rows), the signal reaches its maximum for co- and counter-rotating XUV and NIR pulses. Notably though, for $d$ states with energy above the $np$ state represented by $\mathcal{G}_{d}^{+}$, the signal is maximum for co-rotating pulses, in contrast to $\mathcal{G}_{s}^{+}$ where the counter-rotating pulses maximize the signal. Vice versa, $\mathcal{G}_{d}^{-}$ is maximized for counter-rotating pulses while $\mathcal{G}_s^{-}$ is maximized for co-rotating pulses. This is again a result of selection rules. Assuming that a RCP XUV photon is absorbed from the ground state, leading to the excitation of an $np_+$ state, absorption of a co-rotating RCP NIR photon cannot lead to an excitation of an $s$ state since the final $m=2$ is larger than the angular momentum of the $s$ state $\ell=0$. On the other hand, coupling to a $d_2$ state ($\ell=m=2$) can occur. Hence, the terms that depend on the relative ellipticity of the XUV and NIR field can be seen as a manifestation of circular-dichroism in ATAS. Indeed, if one compares the ATAS signal obtained for a fixed ellipticity of the XUV $\epsilon_X$ for co- and counter-rotating circularly-polarized NIR pulses, the resulting circular dichroism (CD) is given by $\mathrm{CD}(\epsilon_X)=S(\epsilon_X,\epsilon_L=1)-S(\epsilon_X,\epsilon_L=-1)\propto\epsilon_X/(1+\epsilon_X^2)$, as obtained in the previous cDATAS manuscript \cite{Drescher2025aa}.

\begin{figure*}
    \centering
    \includegraphics[width=1.0\linewidth]{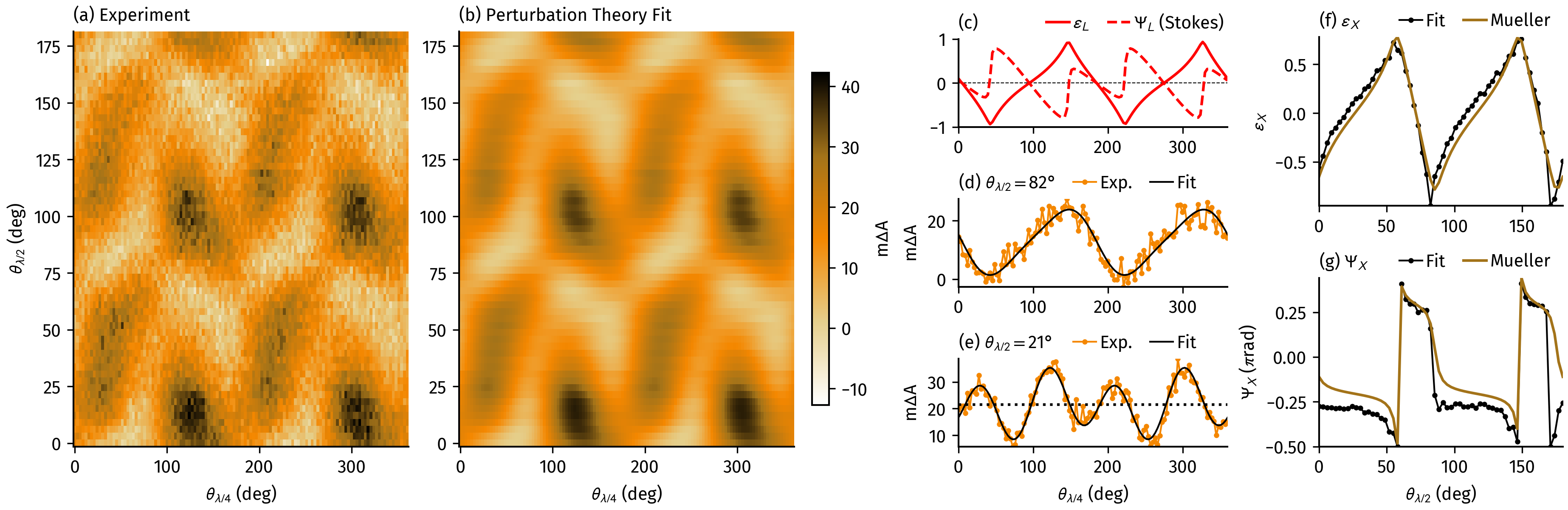}
    \caption{(a) Experimental ATAS signal at the LIS of the 1s2s state at $E_{\mathrm{LIS}}\simeq22.1$ eV. The horizontal and vertical axes correspond to the quarter-wave and half-wave plates, controlling, respectively, the NIR and XUV polarizations. (b) ATAS signal obtained after fitting the $\mathcal{G}_{s}^{-}$ form factor to the experimental signal. (c) The ellipticity and polarization angle of the NIR field after passing through the quarter-wave plate obtained from Mueller's calculus. (d) Experimental (dotted ochre) and perturbation theory (solid black line) circular dichroic ATAS signal at $\theta_{\lambda/2}=82^{\circ}$, corresponding to a predicted $\epsilon_{X}=-0.84$, for varying quarter-wave plate angles. (e) Experimental and perturbation theory linear dichroic ATAS signal at $\theta_{\lambda/2}=21^{\circ}$, corresponding to a predicted $\epsilon_X=0.02$ and $\Psi_X=-0.92$ (dotted line shows the mean over  $\theta_{\lambda/4}$). (f,g) Reconstructed ellipticity (f) and polarization angle (g) of the XUV pulse versus XUV half wave plate angle at the frequency $E_{\mathrm{LIS}}\simeq22.1$ eV (points) and comparison to the ideal values (solid line) of the four-mirror polarizer based on Mueller calculus.}
    \label{fig:Fig3}
\end{figure*}
When the pulses are elliptically polarized $|\epsilon|<1$, terms proportional to the relative angle between the polarization ellipses of the XUV and NIR pulses $\Psi_X-\Psi_L$ contribute as well to the ATAS signal. These terms are not sensitive to the relative ellipticities of the pulses, are null when either of the pulses is circularly polarized $|\epsilon|=1$, are maximized by parallel orientations of the polarization ellipses ($\Delta\Psi=k\pi$, where $k$ is an integer) and minimized by orthogonal orientations ($\Delta\Psi=k\pi/2$). Moreover, note that since the medium is parity-symmetric, the signal is symmetric for $\Delta\Psi$ and $-\Delta\Psi$. Hence, these terms can be seen as a manifestation of lDATAS \cite{Reduzzi:2015aa}. Subtracting the ATAS spectrum for parallel and orthogonal polarizations of XUV and NIR, we obtain a linear dichroism $\mathrm{LD}(\epsilon_X,\epsilon_L)\propto (1-\epsilon^2_X)(1-\epsilon^2_L)/[(1+\epsilon^2_X)(1+\epsilon^2_L)]$, which is maximized when the fields are linearly polarized $\epsilon_X=\epsilon_L=0$. The dependence is again easily understood via selection rules. For $\mathcal{G}_{d}^{\pm}$ only the $m=0$ state contributes to linear dichroism, since it can be coupled to both $np_+$ and $np_-$ states via absorption/emission of one NIR photon. Their contribution is weighted by the coefficient appearing in the form factors as $\mathcal{C}_0$, which arises from the electric-dipole transition matrix element coupling an $np_m$ ($\ell=1$) state to a $d_0$ ($\ell=2$, $m=0$) state. $m=\pm2$ states cannot be coupled to $np$ states with a magnetic quantum number of opposite sign via electric dipole coupling, since this would require a change of magnetic quantum number $\Delta m=\pm2$. Hence, $d_{\pm2}$ states contribute only to circular dichroism but not to linear dichroism. Their contribution in the $\mathcal{G}_d^{\pm}$ form factors are weighted by the coefficient $\mathcal{C}_2$. Since $C_2/C_0=6$, the angle $\Delta\Psi$ dependence of the ATAS signal is largely suppressed compared to the case of $s$ states.   

Given this intuitive understanding of the transient absorption signal dependence on the polarization parameters of the pulses in eDATAS, we  now validate the theory by performing an ATAS experiment in helium using the setup already described in \cite{Drescher2025aa}. Briefly, mJ-level few-cycle NIR pulses are split at a beamsplitter into two arms. Most of the pulse intensity is used in the first arm, which enters a vacuum system where it drives high harmonic generation in krypton to provide a short train of attosecond pulses in the XUV spectral range. After separation of the fundamental driving field from the generated XUV pulses by a thin aluminum foil, the beam passes through a four-mirror polarizer \cite{Schmising:2017aa} before passing through the central hole of an annular mirror. In the second arm, NIR pulses travel for a variable distance to control the temporal delay before also entering the vacuum system and recombining with the first arm co-linearly by reflection off the annular mirror. The combined beams are focused into a gas cell flushed with helium. Afterwards, the XUV beam is spectrally dispersed on a grating and imaged by a CCD camera. A shutter placed in the NIR arm allows to record XUV spectra of the transmission through the gas cell with NIR exposure and without in quick succession to suppress artifacts from fluctuation in the laser parameters. Consequently, the change in absorbance is calculated as: $\Delta A=-\log(I_\textrm{NIR on}/I_\textrm{NIR off})$. Multiple optics in both beams allow for systematic and independent manipulation of the polarization state of the individual beams. In the second arm, an achromatic quarter waveplate mounted on a motorized stage allows one to change the ellipticity of the NIR pulses, while also changing the polarization angle. In the first arm, the four-mirror polarizer, consisting of four molybdenum optics, induces an ellipticity due to the different complex reflectivities of the s- and p-components. While in principle the ellipticity is controllable via the angle of incidence, rotating this reflective polarizer induces a large change in the beam propagation direction. For this reason, a half waveplate is introduced in the first arm before HHG to change the polarization angle of the driving field and thereby the emitted XUV beam. Rotation of this half waveplate therefore controls the XUV ellipticity and the relative polarization angle. While the achromatic waveplates behave close to ideal waveplates, spatial constraints in the construction of the setup require additional optics that degrade the performance of the optics. Both arms include a reflection by a silver mirror under 45$^{\circ}$ angle of incidence directly after the respective waveplate, which adds additional phase-shifts to the beams polarization state if the incident light is not linearly polarized in the s- or p-direction. These effects are included in our analysis using Mueller calculus (see Appendix B).

As done previously in the literature \cite{Drescher2025aa,Reduzzi:2015aa}, we focus on the ATAS signal corresponding to the light induced structure (LIS) produced by the absorption of one NIR photon from the $1s2s$ state at $E_{\mathrm{LIS}}\simeq22.1$ eV, which provides a good benchmark given its strong spectral magnitude compared to neighboring LIS features. As a LIS occurs only when the NIR and XUV pulse overlap in time, we consider the signal at zero relative time-delay between the two. As also pointed out earlier in the text, the form factors $\mathcal{G}^{\alpha}_\ell$ are independent of the time-delay $\tau$ and the XUV frequency. The experimental signal is recorded by varying both the quarter-wave plate angle $\theta_{\lambda/4}$, which allows us to scan over a range of ellipticities $\epsilon_L$ and polarization angles $\Psi_L$ of the NIR field, as well as the half-wave plate angle $\theta_{\lambda/2}$, which allows us to scan over a range of ellipticities $\epsilon_X$ and polarization angles $\Psi_X$ of the XUV field. 
The experimental results are shown in Fig. \ref{fig:Fig3}a. In Fig. \ref{fig:Fig3}c, the predicted NIR ellipticity (solid red line) and polarization angle (dashed red line) dependences on the quarter-waveplate angle $\theta_{\lambda/4}$ in the experiment, as obtained using Mueller's calculus (see Appendix B), are shown. Given that the LIS is associated with an $s-$state with energy below the first member of the $1snp$ Rydberg series at $E_{2p}=21.8$ eV, the ellipticity and polarization angle signal dependence should be well reproduced by the form factor $\mathcal{G}_s^{-}$ in Eq. \ref{eq:formfactor}. We then take as a fit function for the experimental results at a given quarter-wave plate angle $\theta_{\lambda/4}$ the function $f(\epsilon_X,\Psi_X)=\mathcal{A}\mathcal{G}_s^{-}(\epsilon_X,\Psi_X)+\mathcal{B}$, where the independent variables are the ellipticity $\epsilon_L$ and polarization angle $\Psi_L$ of the NIR field as a function of $\theta_{\lambda/4}$ (as shown in Fig. \ref{fig:Fig3}c), and the fit parameters are the XUV ellipticity $\epsilon_X$ and polarization angle $\Psi_X$, as well as the constants $\mathcal{A}$ and $\mathcal{B}$. Fig. \ref{fig:Fig3}b shows the resulting map obtained from the fit, which is in excellent agreement with the experimental results. 

To confirm the correctness of our fit, we focus on the ATAS signal recorded at the half-waveplate angles where the fitted ellipticity of the XUV pulse is at its maximum at $\theta_{\lambda/2}=82^{\circ}$, and where the XUV is linear at $\theta_{\lambda/2}=21^{\circ}$. At $\theta_{\lambda/2}=82^{\circ}$, where we obtain $\epsilon_X\simeq-0.84$, the corresponding experimental (dotted line) and fitted (solid line) signal are shown in Fig. \ref{fig:Fig3}d. As predicted from the $\mathcal{G}_{s}^{-}$ form factor, the dichroic signal is maximized for co-rotating and minimized by counter-rotating NIR and XUV pulses. At the half-waveplate angle $\theta_{\lambda/2}=21^{\circ}$, where the fitted XUV values are $\epsilon_X\simeq 0.02$ and $\Psi_X\simeq-0.92$, the experimental and fit results are shown in Fig. \ref{fig:Fig3}e). 
In this case, at the quarter-wave plate positions where the NIR is highly elliptical ($\epsilon_L$ approaches unity), the experimental signal is at half of the difference between the maximum and minimum value of the scan over the quarter-wave plate angle (dotted line). The modulation of the linear dichroic signal, including four local maxima and four local minima, arises due to the varying relative angle between the XUV and NIR and is reproduced by the linear dichroic term $\cos(2\Delta\Psi)$ (not shown in Fig. \ref{fig:Fig3}e). Thus, we can conclude that the experiment and theory are in excellent agreement, and that both cDATAS and lDATAS should be seen as limiting cases of the more general eDATAS technique.
From an experimentalist point of view, it is worth pointing out that the signal for a highly elliptical XUV pulse shows four extrema over the whole rotation of the $\lambda/4$ waveplate, while it shows eight extrema for a linearly polarized XUV pulse. This provides a quick optimization pathway of the XUV polarizer if either the most elliptical or linear polarization is desired.

Given the excellent agreement between theory and experiment, we can also use eDATAS to reconstruct the polarization state of the XUV field at a given frequency. Taking again the LIS associated to the $2s$ state at $E_{\mathrm{LIS}}=22.1$ eV as an example, the reconstructed XUV ellipticity and polarization angle for varying angle of the half-wave plate is shown in the panels (f) and (g) in Fig. \ref{fig:Fig3}, compared to the result for an ideal four-mirror polarizer based on Mueller calculus (see Appendix B). We note that this is a task we accomplish here self-consistently, without dedicated XUV polarimetry setups \cite{Bengs:2021aa}, relying instead on the much easier measurement of the polarization state of an NIR pulse. In principle, the same procedure can be repeated at other frequencies, allowing polarimetry of the XUV pulse across its entire spectral bandwidth. In spectral regions where contributions from several $ns$ or $nd$ states overlap, however, the fit function must include all the corresponding form factors $\mathcal{G}_\ell^{\alpha}$, each with a spectrally dependent weight. A full polarimetric characterization of the attosecond pulse lies beyond the scope of this work, but we expect that accurate TDSE simulations, combined with an independent reconstruction of the spectral phase and intensity of both pulses — for instance via standard XUV polarimetry \cite{Bengs:2021aa} used as a one-time benchmark — would establish eDATAS as a reliable, self-contained technique for XUV polarimetry. 

Having obtained the parameters $\epsilon_X$, $\epsilon_L$, $\Psi_X$ and $\Psi_L$ for each measurement, as well as the expressions for the form factors, we can now identify the character of each LIS due to their dynamical dependence on $\Delta\Psi=\Psi_X-\Psi_L$ and $\epsilon_X\cdot\epsilon_L$: For this, we sort and bin the experimental data at each XUV frequency into a three-dimensional matrix, $A_{X,Y,Z}(\omega)$, where $X$ and $Y$ take three bins each over the XUV and NIR ellipticity between $(-1,1)$, respectively, and $Z$ takes 14 bins for $0\leq\Delta\Psi\leq\tfrac{\pi}{2}$.

\begin{figure}
    \centering
    \includegraphics[width=\linewidth]{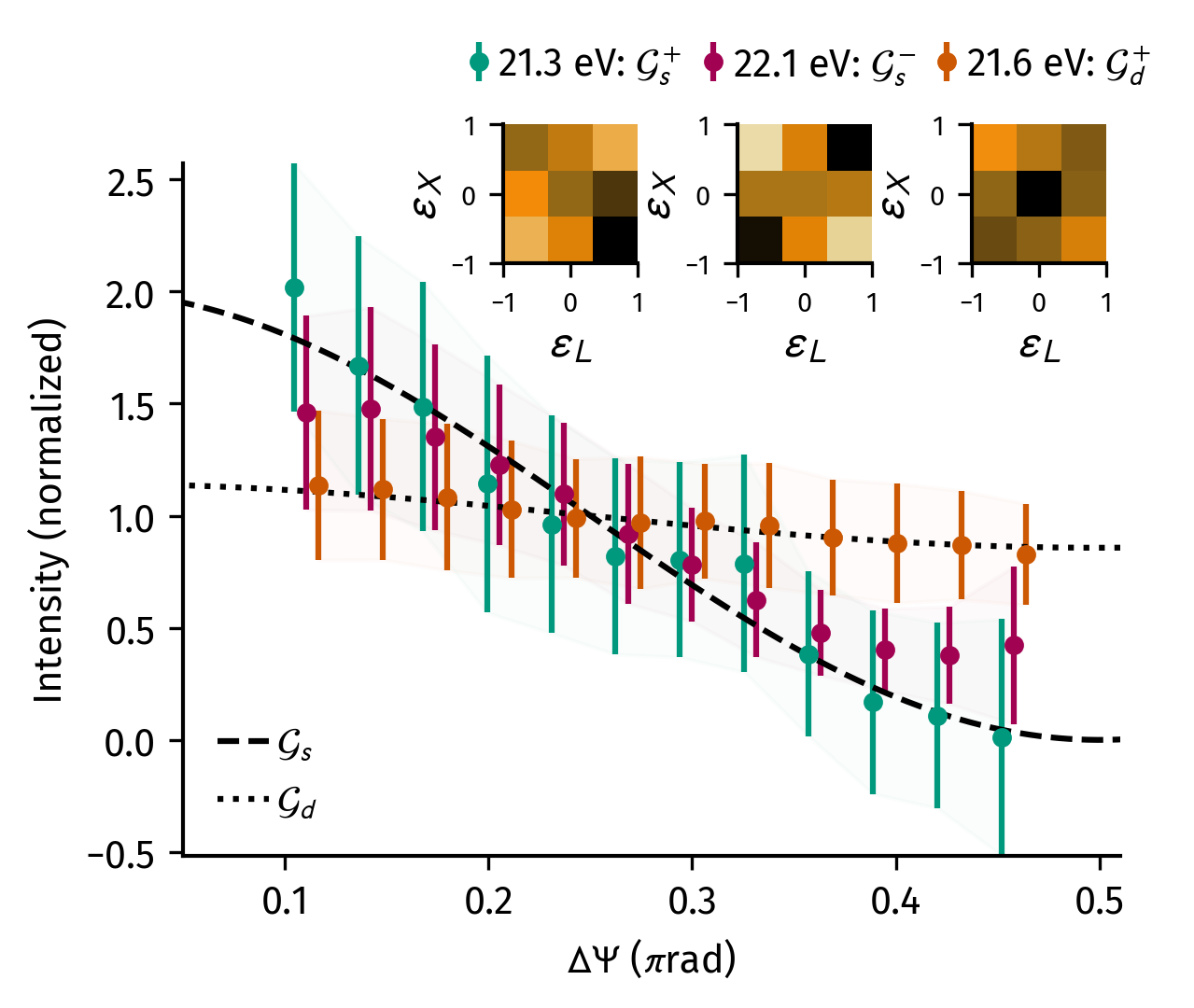}
    \caption{Dependence on the ellipsis angle ($\Delta\Psi$) of selected lines, allowing their comparison to the $\mathcal G_s$ and $\mathcal G_d$ behavior. The experimental data has been binned according to the XUV and NIR ellipticities and ellipsis angle. Bars indicate the standard deviation at each bin. The points have been horizontally offset to ease visualization. Insets: The helicity dependence of each line. Together with the $\Delta\Psi$-dependence they allow to identify the form factors (cf Fig.~\ref{fig:Fig2}) of each feature.}
    \label{fig:Fig4}
\end{figure}

To identify the character, the data are first plotted over $\Delta\Psi$ for the $\epsilon_X\sim\epsilon_L\sim0$ bin. The results can be seen in the main plot of Fig. \ref{fig:Fig4} for three exemplary photon energies, 21.3\,eV, 21.6\,eV and 22.1\,eV. While the data at 21.3\,eV and 22.1\,eV show a strong dynamic with $\Delta\Psi$, the data at 21.6\,eV show a strongly suppressed dynamic. Comparison to the behaviors of $\mathcal{G}_s^\pm$ (strong dynamic) and $\mathcal{G}_d^\pm$ (suppressed dynamic) allows to identify that the signals at 21.3\,eV and 22.1\,eV originate from couplings with $s$-states and the signal at 21.6\,eV originates from couplings with $d$-states. We then compare the signal at each bin for $\epsilon_X$ and $\epsilon_L$, averaged over all $\Delta\Psi$. The results are shown as color maps in the insets of Fig. \ref{fig:Fig4}. As can be seen, the signals at 22.1\,eV and 21.6\,eV show a strong response at the diagonal ($\epsilon_X\cdot\epsilon_L>0$) while the signal at 21.3\,eV shows a strong response at the anti-diagonal ($\epsilon_X\cdot\epsilon_L<0$). Comparing with Fig.~\ref{fig:Fig2} allows us to complete the assignment of the LIS: The feature at 21.3\,eV follows $\mathcal{G}_s^+$, the feature at 22.1\,eV follows $\mathcal{G}_s^-$ and the feature at 21.6\,eV follows $\mathcal{G}_d^+$. Knowledge of their form factors, their XUV energy and the NIR energy ($\sim$1.6\,eV) then finally allows to identify their origin: The feature at 21.3\,eV is a LIS of the $3s$ dark state (22.9\,eV) that is coupled to via the lower lying $2p$ state. The feature at 22.1\,eV is the previously identified LIS of the $2s$ dark state (20.6\,eV), coupled to via the higher lying $2p$ state. Lastly, the feature at 21.6\,eV is an LIS of the $3d$ dark-state near 23.1\,eV, coupled via the lower lying $2p$ state.

Note that while we have concentrated our analysis on the LIS, the same analysis can be applied to the light induced phase, i.e. the reshaping of the bright-state lineshape due to NIR-driven Rabi cycling~\cite{Wu:2016aa,Drescher:2020aa}, at least while the spectral reshaping remains linear with the light-induced phase~\cite{Drescher:2019aa}. Moreover, we note that higher-order multiphoton processes could also become important at higher NIR intensities, inducing LIS arising either from the coupling of two $np$ states via intermediate $ns$ or $nd$ states, or from $nf$ states coupled to $np$ states via two-photon transitions. In that case, the ellipticity dependence can be used to identify these LIS. The $np\rightarrow n'p$ features vanishes for both co- and counter-rotating circularly polarized pulses, thus requiring elliptically polarized fields to appear at all, in contrast to the one-photon form factors of Eqs. (\ref{eq:formfactor}). Features involving $nf$ dark states, on the other hand, are maximal for co-rotating circular pulses; they can be nonetheless distinguished by their NIR ellipticity dependence, which involves higher powers of $\epsilon_L$ than the one-photon case: a two-photon NIR process yields a form factor of degree 4 in $\epsilon_L$ over a normalization $(1+\epsilon_L^2)^2$, so that no combination of the $\mathcal{G}_\ell^{\alpha}$ of Eqs. (\ref{eq:formfactor}), which are quadratic in $\epsilon_L$, can reproduce it. Nonetheless, given the low NIR intensities used in this work, these features are negligible in the experiment.

In conclusion, our work establishes eDATAS as a new addition to the toolkit of attosecond spectroscopy techniques. Within this framework, cDATAS and lDATAS emerge as limiting cases: cDATAS for co-/counter-rotating circularly polarized pulses, and lDATAS for parallel/orthogonal linearly polarized XUV and NIR pulses. Here we have focused on the eDATAS spectrum of an isotropic gas of atoms, resulting in signal that is invariant under reflections, i.e., $\mathcal{G}_\ell^\alpha(\Delta\Psi)=\mathcal{G}_\ell^{\alpha}(-\Delta\Psi)$. From a two-pathway interference perspective, this symmetry arises because the two interferometric ‘arms’ associated with the $p_m$ states contribute with equal amplitude but opposite phase to the total dipole moment. In contrast, this symmetry can be broken in systems such as open-shell atoms, aligned or oriented molecular ensembles, and chiral \cite{Drescher:2020aa} or spin–orbit-coupled media. The resulting eDATAS signal should then retain reflection-odd terms proportional to $\sin(2\Delta\Psi)$. eDATAS therefore holds potential as a sensitive probe of symmetry breaking phenomena triggered by electron motion on the attosecond timescale, measured by tracking the phase between the reflection-odd and reflection-even terms as a function of the XUV–NIR delay. The relative phase between these terms can contain valuable information on the phase of the involved transitions, allowing one to time the buildup of the two contributions to the total signal. We also note that it is possible to extend eDATAS beyond the dipole approximation and include magnetic and higher multipole interactions between the light wave and the system under study. Including magnetic interactions could, for example, unlock the potential of eDATAS as an ultrafast magnetometry technique, in which transient magnetic fields are reconstructed by tracking time-dependent Zeeman shifts in the ATAS spectrum \cite{Hu:2025aa}, and spin dynamics are isolated by comparing reflection-odd and reflection-even contributions. Moreover, we note that the combination of eDATAS with structured light \cite{Rubinsztein:2017aa,Angelsky:2020aa,Forbes:2021aa} could further extend the technique into the richer realm of topological light–matter interactions \cite{Mayer:2024aa,Garcia:2024aa}, where the spatial structure of the field becomes an additional control knob for the attosecond dynamics.

\section*{Acknowledgments}

Investigations were supported by the U.S. Air Force Office of Scientific Research, Grant Nos FA9550-24-1-0184 and FA9550-19-1-0314. LBD acknowledges the European Union's Horizon research and innovation programme under the Marie Sklodowska-Curie grant agreement No 101066334 — SR-XTRS-2DLayMat. This work was funded by UK Research and Innovation (UKRI) under the UK government’s Horizon Europe funding guarantee [grant number EP/Z001390/1]. NM gratefully acknowledges helpful discussions with Margarita Khokhlova and Emilio Pisanty. \\
NM dedicates this work to the memory of Marina Mayer.

\section*{Appendix A: Perturbation theory derivation of $\mathcal{G}_{\ell}^{\pm}$}

Here we derive the transient absorption signal of a hydrogenic atom in the presence of elliptically polarized XUV and NIR pulses, with an arbitrary relative angle between the polarization ellipses. Our aim is to solve the TDSE $\text{i}\partial_t|\Psi(t)\rangle=\hat{H}(t)|\Psi(t)\rangle$, where the Hamiltonian in the length gauge and dipole approximation is given by $\hat{H}(t)=\hat{H}_0+\hat{\mathbf{r}}\cdot\mathbf{E}(t)$. Here $\hat{H}_0$ is the atomic Hamiltonian and $\hat{\mathbf{r}}$ is the position operator. The two-color electric field is given by $\mathbf{E}(t)=\mathbf{E}_X(t)+\mathbf{E}_L(t-\tau)$, where the subscript $X/L$ indicate, respectively, the XUV and NIR fields, $\tau$ is the relative delay between the fields, and the general form for the electric field in the circular basis is given in Eq. \ref{eq:fields}. As an ansatz for the TDSE we use an expansion over the bound field-free states of the atomic Hamiltonian $|\Psi(t)\rangle=\sum_i c_i(t)e^{-\text{i}(E_i-\text{i}\Gamma_i/2)t}|i\rangle$, where $E_i$ is the field-free energy of the atomic state ($E_i=0$ for the ground state) and $\Gamma_i=1/\tau_i$, where $\tau_i$ is the spontaneous decay lifetime of the field-free state. From now on we assume for simplicity a common width $\Gamma_i=\Gamma$ for all atomic states except the ground state $\Gamma_g=0$, a good approximation given the relevant timescales in the experiment. We also exclude continuum states and focus on the response of the bound part of the atomic spectrum.

Rabi frequencies in the Rotating-Wave Approximation (RWA) are denoted as
\begin{equation}
    \label{eq:Rabi}\Omega^{i/f}_{Q,\pm}=\frac{\mathcal{E}_Q}{\sqrt{2(1+\epsilon_Q^2)}}\frac{(1\pm\epsilon_Q)}{2}e^{\pm\text{i}\Psi_Q}\mathbf{r}^{i/f}\cdot\mathbf{e}_{\mp},
\end{equation}
where $\mathbf{r}^{i/f}=\langle f|\hat{\mathbf{r}}|i\rangle$, the subscript $Q=\{X,L\}$ indicates whether a transition is driven by the XUV or NIR field respectively. The transition matrix element for hydrogenic states $\langle\mathbf{r}|k\rangle=R_k(r)Y_{l_km_k}(\Omega)$ is given by
\begin{equation}
    \langle f|\hat{\mathbf{r}}|i\rangle=\sqrt{\frac{4\pi}{3}}\sum_{q=-1}^{1}\langle f|rY_{1q}|i\rangle\mathbf{e}^*_q,
\end{equation}
where $Y_{lm}$ is a spherical harmonic, $\Omega$ is the solid angle and $\mathbf{e}_{\pm}$ are defined as in the main text above and $\mathbf{e}_0=\mathbf{e}_z$. The selection rules for one photon electric dipole transitions couple states with opposite parity $|\ell_f-\ell_i|=1$, where absorption of a RCP ($\mathbf{e}_-$) or LCP ($\mathbf{e}_+$) photon leads to a final state with magnetic quantum number $m_f=m_i\pm1$ respectively. 

Let us then consider as a starting example the case of two magnetic sublevels $|np_\pm\rangle=|n,\ell=1,m=\pm1\rangle$ coupled to the ground state via the attosecond pulse, and to a neighboring $|ks\rangle=|k,\ell=0,m=0\rangle$ state with energy $E_{ks}<E_{np}$ via the NIR pulse. The corresponding amplitudes are governed by the following coupled differential equations
\begin{widetext}
    \begin{eqnarray}
    \text{i}\dot{c}_g(t)&=&\sum_m\Omega_{X,-m}^{g/np_m}f_X(t)c_{np_m}(t)e^{-\text{i}(E_{np}-\omega_X)t}e^{-\frac{\Gamma}{2}t}\\
    \text{i}\dot{c}_{np_m}(t)&=&\Omega_{X,m}^{np_m/g}f_X(t)c_g(t)e^{\text{i}(E_{np}-\omega_X)t}e^{\frac{\Gamma}{2}t}+\Omega_{L,-m}^{ks/np_m}f_L(t-\tau)c_{ks}(t)e^{\text{i}(E_{np}-E_{ks}-\omega_L)(t-\tau)}\\
    \text{i}\dot{c}_{ks}(t)&=&\sum_m\Omega_{L,m}^{np_m/ks}f_L(t-\tau)c_{np_m}(t)e^{-\text{i}(E_{np}-E_{ks}-\omega_L)(t-\tau)},
\end{eqnarray}
\end{widetext}

where $\tau$ is the time-delay between the peak of the XUV and NIR pulses and $\Omega_{L,-m}^{np_m/ks}=(\Omega_{L,m}^{ks/np_m})^*$. Assuming $\tau>0$ (the NIR pulse arrives after the XUV) and modeling a weak attosecond pulse $\mathcal{E}_X\ll1$ as a $\delta-$like impulse $f_X(t)=\delta(t)$, we can assume the ground state is not affected $c_g(t)\simeq1$ and reduce the system of coupled differential equations to the excited amplitudes
\begin{widetext}
\begin{eqnarray}
    \text{i}\dot{c}_{np_m}(t)&=&\Omega_{X,m}^{np_m/g}\delta(t)e^{\text{i}(E_{np}-\omega_X)t}e^{\frac{\Gamma}{2}t}+\Omega_{L,-m}^{ks/np_m}f_L(t-\tau)c_{ks}(t)e^{\text{i}(E_{np}-E_{ks}-\omega_L)(t-\tau)}\\
    \text{i}\dot{c}_{ks}(t)&=&\sum_m\Omega_{L,m}^{np_m/ks}f_L(t-\tau)c_{np_m}(t)e^{-\text{i}(E_{np}-E_{ks}-\omega_L)(t-\tau)}.
\end{eqnarray}
\end{widetext}

We now solve this system of equations using perturbation theory with respect to the perturbing NIR field $c_{np_m}(t)=c_{np_m}^{(0)}(t)+c_{np_m}^{(1)}(t)$, where the zero-order amplitude corresponds to the initial excitation due to the XUV
\begin{equation}
   c_{np_m}^{(0)}(t)=-\text{i}\Theta(t)\Omega_{X,m}^{np_m/g}.
\end{equation}
Inserting the zero-order amplitude in the equation for the $c_{ks}$ amplitude we obtain the formal solution for $t>0$
\begin{eqnarray}
    c_{ks}(t)&=&-\sum_m\Omega_{X,m}^{np_m/g}\Omega_{L,m}^{np_m/ks}\nonumber\\
    &&\int_0^{t}dt'f_L(t'-\tau)e^{-\text{i}(E_{np}-E_{ks}-\omega_L)(t'-\tau)}.
\end{eqnarray}
Substituting back into the equation for the $c_{np_m}(t)$ amplitudes we obtain the formal solution $c_{np_m}(t)=c_{np_m}^{(0)}(t)+c_{np_m}^{(1)}(t)$, with
\begin{eqnarray}
    c_{np_m}(t)&=&-\text{i}\Theta(t)\left[\Omega_{X,m}^{np_m/g}-\right.\nonumber\\
    &&\left.\Omega_{L,-m}^{ks/np_m}\sum_{m'}\Omega_{X,m'}^{np_{m'}/g}\Omega_{L,m'}^{np_{m'}/ks}G_L(t,\tau)\right].
\end{eqnarray}
where $G_L(t,\tau)=\int_{0}^{t}dt''\int_{0}^{t''}dt'f_L(t''-\tau)f_L(t'-\tau)e^{-\text{i}(E_{np}-E_{ks}-\omega_L)(t'-t'')}$. The positive frequency part of the induced time-dependent dipole moment between ground and excited states is then
\begin{equation}
    \mathbf{d}(t)=\sum_{m}c_{np_m}(t)e^{-\text{i}E_{np}t-\frac{\Gamma}{2}t}\langle g|\mathbf{r}|np_m\rangle
\end{equation}
and the circular components of its Fourier transform $\tilde{\mathbf{d}}(\omega)=\int e^{\text{i}\omega t}\mathbf{d}(t)dt=\sum_m\tilde{\mathbf{d}}_m(\omega)$ can be written as $\tilde{\mathbf{d}}_m(\omega)=\tilde{\mathbf{d}}_m^{(0)}(\omega)+\tilde{\mathbf{d}}_m^{(1)}(\omega)$, where
\begin{eqnarray}
    \tilde{\mathbf{d}}_m^{(0)}(\omega)&=&-\text{i}A_{\Gamma}(\omega)\Omega_{X,m}^{np_m/g}\mathbf{r}^{np_m/g}\\
    \tilde{\mathbf{d}}_m^{(1)}(\omega)&=&\text{i}\tilde{G}_L(\omega,\tau)\mathbf{r}^{np_m/g} \Omega_{L,-m}^{ks/np_m}\nonumber\\
    &&\times\sum_{m'}\Omega_{X,m'}^{np_{m'}/g}\Omega_{L,m'}^{np_{m'}/ks}
\end{eqnarray}
where $A_\Gamma(\omega)=\frac{1}{\frac{\Gamma}{2}-\text{i}(\omega-E_{np})}$ is the lineshape profile and for NIR pulses shorter than the natural decay of the excited states
\begin{eqnarray}
    \tilde{G}_L(\omega,\tau)&=&\int_0^\infty dt\, e^{\text{i}(\omega-E_{np})t}e^{-\frac{\Gamma}{2}t}G_L(t,\tau)\\
    &\simeq&\frac{1}{2}e^{\text{i}(\omega-E_{np})\tau}e^{-\frac{\Gamma}{2}\tau}\nonumber\\
    &&\frac{\tilde{f}_L(\omega-E_{ks}-\omega_L)\tilde{f}_L(E_{ks}+\omega_L-E_{np})}{\frac{\Gamma}{2}-\text{i}(\omega-E_{np})},\nonumber
\end{eqnarray}
Finally, the ATAS response $S(\omega)=\omega\mathrm{Im}[\tilde{\mathbf{d}}(\omega)\cdot\tilde{\mathbf{E}}^*_X(\omega)]$ can be written as
\begin{widetext}
\begin{eqnarray}
    &&S(\omega)=S^{(0)}(\omega)+S^{(1)}(\omega)\\
    &&S^{(0)}(\omega)=\mathcal{A}^{(0)}\frac{\omega \Gamma}{\left(\frac{\Gamma}{2}\right)^2+(\omega-E_{np})^2}\\
    &&S^{(1)}(\omega)=-\mathcal{A}^{(1)}\frac{\omega\mathcal{H}(\omega,\tau)}{\left(\frac{\Gamma}{2}\right)^2+(\omega-E_{np})^2}\mathcal{G}(\epsilon_X,\epsilon_L,\Psi_X-\Psi_L)\\
    &&\mathcal{G}(\epsilon_X,\epsilon_L,\Psi_X-\Psi_L)=|\mathbf{r}^{np/ks}|^2\frac{\left[(1+\epsilon_X)^2(1+\epsilon_L)^2+(1-\epsilon_X)^2(1-\epsilon_L)^2+2(1-\epsilon_L^2)(1-\epsilon_X^2)\cos(2(\Psi_X-\Psi_L))\right]}{(1+\epsilon^2_X)(1+\epsilon^2_L)}
\end{eqnarray}
\end{widetext}
where $\mathcal{A}^{(0)}=\mathcal{E}^2_X|\mathbf{r}^{np/g}|^2/2^3$, $\mathcal{A}^{(1)}=\mathcal{E}^2_X\mathcal{E}^2_L|\mathbf{r}^{np/g}|^2/2^{6}$ and for a Gaussian NIR pulse $f_L(t)=\exp[-(t/\sigma_L)^2]$
\begin{eqnarray}
    \mathcal{H}(\omega,\tau)&=&\frac{\pi\sigma^2_L}{4}e^{-\sigma_L^2(\omega-E_{\mathrm{LIS}})^2/4}e^{-\sigma_L^2\Delta^2/4}e^{-\frac{\Gamma}{2}\tau}\nonumber\\
    &&\times\left(\Gamma\cos[(\omega-E_{np})\tau]-\right.\nonumber\\
    &&\left.2(\omega-E_{np})\sin[(\omega-E_{np})\tau]\right).
\end{eqnarray}
We see that the absorption line described by $S^{(0)}(\omega)$ is modified by a term that depends both on the position of the light-induced state $E_{\mathrm{LIS}}=E_{ks}+\omega_L$, as well as on the detuning between the NIR laser frequency and the transition frequency $\Delta=E_{np}-E_{ks}-\omega_L$. The delay dependence indicates the reshaping of the ATAS signal between Lorentzian and Fano-like lineshapes. Most relevant for this work, the modification to the ATAS response depends on the relative ellipticity of the XUV and NIR pulses and the relative angle between their polarization ellipses through the term $\mathcal{G}(\epsilon_X,\epsilon_L,\Psi_X-\Psi_L)$. 
If the dark state has energy above the $E_{np}$ line, the corresponding energy of the light-induced state and transition frequency need to be modified by $E_k$, $E_{\mathrm{LIS}}=E_{k}-\omega_L$, $\Delta=E_{np}-E_{k}+\omega_L$, as well as the magnitude of the transition matrix element $|\mathbf{r}^{ks/np_m}|^2$. The ellipticity and polarization angle dependence will also change in accordance with the one-photon transition selection rules. Moreover, if the dark state is a $d$-state with angular momentum $\ell=2$, we need to take into account all the relevant sublevels with magnetic number $m\in[-2,2]$. In order to classify all these cases, we use the notation $\mathcal{G}_{\ell}^{\alpha}$ for the ellipticity and polarization angle dependent factor, where $\ell=\{s,d\}$ indicates the angular momentum of the dark state and $\alpha=\{+,-\}$ indicates whether the dark state energy is above ($+$) or below ($-$) the $p$ state. Following analogous derivations as above, we obtain  
\begin{widetext}
\begin{eqnarray}
    \mathcal{G}_{s}^{-}&=&|\mathbf{r}^{s/p_m}|^2\frac{(1+\epsilon_X)^2(1+\epsilon_L)^2+(1-\epsilon_X)^2(1-\epsilon_L)^2+2(1-\epsilon_L^2)(1-\epsilon_X^2)\cos(2(\Psi_X-\Psi_L))}{(1+\epsilon^2_X)(1+\epsilon^2_L)},\\
    \mathcal{G}_{s}^+&=&|\mathbf{r}^{s/p_m}|^2\frac{(1+\epsilon_X)^2(1-\epsilon_L)^2+(1-\epsilon_X)^2(1+\epsilon_L)^2+2(1-\epsilon_L^2)(1-\epsilon_X^2)\cos(2(\Psi_X-\Psi_L))}{(1+\epsilon^2_X)(1+\epsilon^2_L)},
\end{eqnarray}
\end{widetext}
\begin{widetext}
\begin{eqnarray}
    \mathcal{G}_{d}^-&=&|\mathbf{r}^{d_0/np_m}|^2\frac{\left[(1+\epsilon_X)^2(1+\epsilon_L)^2+(1-\epsilon_X)^2(1-\epsilon_L)^2+2(1-\epsilon_L^2)(1-\epsilon_X^2)\cos(2(\Psi_X-\Psi_L))\right]}{(1+\epsilon^2_X)(1+\epsilon^2_L)}+\nonumber\\
    &&|\mathbf{r}^{d_2/np_m}|^2\frac{\left[(1+\epsilon_X)^2(1-\epsilon_L)^2+(1-\epsilon_X)^2(1+\epsilon_L)^2\right]}{(1+\epsilon^2_X)(1+\epsilon^2_L)},\\
    \mathcal{G}_{d}^+&=&|\mathbf{r}^{d_0/np_m}|^2\frac{\left[(1+\epsilon_X)^2(1-\epsilon_L)^2+(1-\epsilon_X)^2(1+\epsilon_L)^2+2(1-\epsilon_L^2)(1-\epsilon_X^2)\cos(2(\Psi_X-\Psi_L))\right]}{(1+\epsilon^2_X)(1+\epsilon^2_L)}+\nonumber\\
    &&|\mathbf{r}^{d_2/np_m}|^2\frac{\left[(1+\epsilon_X)^2(1+\epsilon_L)^2+(1-\epsilon_X)^2(1-\epsilon_L)^2\right]}{(1+\epsilon^2_X)(1+\epsilon^2_L)}.
\end{eqnarray}
\end{widetext}
Note the different weightings of the contributions for the $d$-state from the $m=\pm2$ and $m=0$ states to the response for the $d$ state, where after factorization of the transition matrix element in terms of angular and radial parts we obtain the expressions reported in the main text. The term that depends on the angle between the polarization ellipses is always due to the $m=0$ state, which couples the sublevels $np_+$ and $np_-$ through the counter-rotating NIR photons. Hence, the total modification to the spectrum around a given $np$ state is a sum over all these contributions, where each contribution depends on the laser frequency $\omega_L$, the specific energy of interest $\omega$ in the XUV absorption spectrum and the magnitude of the transition matrix elements, as well as on the ellipticities $\epsilon$ and polarization angles $\Psi$. 

\section{Appendix B: Mueller's calculus}

The ellipticity and polarization angle of the NIR beam in the experiment and shown in Fig. \ref{fig:Fig3}c are calculated using Mueller's calculus. The Stokes vector $S=(S_0,S_1,S_2,S_3)^{T}$ for a linearly polarized beam at angle $\alpha$ from the horizontal $x$-axis of the laboratory frame is
\begin{equation}
    S(\alpha)=\left(\begin{matrix}1 \\ \cos2\alpha \\ -\sin2\alpha \\ 0 \end{matrix}\right),
\end{equation}
where a s-polarized beam corresponds to $S(0)$ and a p-polarized beam to $S(\pi/2)$. The Mueller's matrices for a reflective surface with phase shift (due to the incidence at 45°) and a general linear retarder (waveplate) are given by
\begin{widetext}
\begin{eqnarray}
    M_{\text{Ag}}&=&\left(\begin{matrix} A & B & 0 & 0 \\
    B & A & 0 & 0 \\
    0 & 0 & C\cos\Delta & C\sin\Delta \\
    0 & 0 & -C\sin\Delta & C\cos\Delta
    \end{matrix}\right)\\
    M_{\text{WP}}(\theta,\delta)&=&\left(\begin{matrix} 1 & 0 & 0 & 0 \\
    0 & \cos(2\theta)^2+\sin(2\theta)^2\cos\delta & \cos(2\theta)\sin(2\theta)(1-\cos\delta) & \sin(2\theta)\sin\delta \\
    0 & \cos(2\theta)\sin(2\theta)(1-\cos\delta) & \cos(2\theta)^2\cos\delta+\sin(2\theta)^2 & -\cos(2\theta)\sin\delta \\ 
    0 & -\sin(2\theta)\sin\delta & \cos(2\theta)\sin\delta & \cos\delta
    \end{matrix}\right)
\end{eqnarray}
\end{widetext}
where $A=(r_s+r_p)/2$, $B=(r_s-r_p)/2$,$C=\sqrt{r_sr_p}$ and for the silver mirror in the experiment, $\Delta=\phi_p-\phi_s$ and $r_s=0.99378$, $r_p=0.99688$, $\phi_{s}=30.724^{\circ}$, $\phi_p=-164.648^{\circ}$~\cite{Johnson:1972aa}. Here $\theta$ is the angle of incidence of a linearly polarized field with respect to the waveplate axis and $\delta$ is the phase delay between the orthogonal components induced by the waveplate. In the experiment, the NIR beam passes through a quarter wave-plate and is then reflected by the silver mirror. The Stokes vector of the NIR beam in the experiment is then given by
\begin{equation}
    S(\theta_{\lambda/4})=M_{\text{Ag}}\cdot M_{\text{WP}}(\theta_{\lambda/4}+\pi/2) \cdot  S(0).
\end{equation}
The ellipticity and polarization angle of the NIR beam are then obtained as
\begin{eqnarray}
    \epsilon_L(\theta_{\lambda/4})&=&\tan\left[\frac{1}{2}\arcsin\left(\frac{S_{3}(\theta_{\lambda/4})}{S_0(\theta_{\lambda/4})}\right)\right],\\
    \Psi_L(\theta_{\lambda/4})&=&\frac{1}{2}\arctan\left[\frac{S_2(\theta_{\lambda/4})}{S_1(\theta_{\lambda/4})}\right],
\end{eqnarray}
or more explicitly,
\begin{widetext}
\begin{eqnarray}
    \epsilon_L(\theta)&=&\tan\left[\frac{1}{2}\arcsin\left(\frac{C\sin(2\theta)(\cos\Delta-\sin\Delta\cos(2\theta))}{A+B\cos^2(2\theta)}\right)\right],\\
    \Psi_L(\theta)&=&\frac{1}{2}\arctan\left[\frac{C\sin(2\theta)(\cos\Delta\cos(2\theta)+\sin\Delta)}{B+A\cos^2(2\theta)}\right].
\end{eqnarray}
\end{widetext}
For the XUV, the Stokes vector is given by
\begin{widetext}
\begin{eqnarray}
S(\theta_{\lambda/2},\omega_X)=R^{-1}(\alpha)\left[\mathrm{M}_{\mathrm{Mo}}(\omega_X)\right]^4R(\alpha)\mathrm{M}_{\mathrm{WP}}(\theta_{\lambda/2},\pi)S(\pi/2),
\end{eqnarray}
\end{widetext}
where the Molybdenum polarizer optics are treated as a linear retarder whose Mueller matrix at photon energy $\omega_X$ is given by
\begin{widetext}
\begin{eqnarray}
    \mathrm{M}_{\mathrm{Mo}}(\omega_X)&=&\frac{1}{2}(r_p^2+r_s^2)\left(\begin{matrix}1 & -\cos2\psi & 0 & 0 \\ -\cos2\psi & 1 & 0 & 0 \\
    0 & 0 & \sin2\psi\cos\Delta & \sin2\psi\sin\Delta \\ 0 & 0 & -\sin2\psi\sin\Delta & \sin2\psi\cos\Delta\end{matrix}\right),
\end{eqnarray}
\end{widetext}
where $\tan(\psi(\omega_X))=r_p(\omega_X)/r_s(\omega_X)$ and $\Delta(\omega_X)=\phi_p(\omega_X)-\phi_s(\omega_X)$. Here $\phi_{s/p}(\omega_X)$ and $r_{s/p}(\omega_X)$ are the phase and moduli of the complex Fresnel reflection coefficients of Molybdenum evaluated at a fixed angle of incidence of $\simeq78^{\circ}$ and taken from tabulated optical constants~\cite{pauly:2020aa}. The polarizer is oriented at an angle $\alpha$ relative to the laboratory frame, accounted for by the rotation matrix
\begin{eqnarray}
R(\alpha)=\left(\begin{matrix}1 & 0 & 0 & 0 \\ 0 & \cos2\alpha & \sin2\alpha & 0 \\ 0 & -\sin2\alpha & \cos2\alpha & 0 \\ 0 & 0 & 0 & 1\end{matrix}\right),
\end{eqnarray}
so that the four successive reflections act in the polarizer frame.

\bibliographystyle{unsrtnat}
\bibliography{biblio}

\end{document}